\documentstyle[aps,multicol,epsfig]{revtex}

\begin{document}

\input{epsf.tex}
\epsfverbosetrue

\title{Bifurcation of gap solitons through catastrophe theory}
\author{Claudio Conti}
\address{Dipartimento di Ingegneria Elettronica,
Universita' di Roma Tre, Via della Vasca
Navale 84, 00146 Roma, Italy}
\author{Stefano Trillo}
\address{Department of Engineering,
University of Ferrara, Via Saragat 1, 44100 Ferrara, Italy\\
and Istitituto Nazionale di Fisica della Materia,
INFM-RM3, Via della Vasca
Navale 84, 00146 Roma, Italy}

\date{\today}

\maketitle

\begin{abstract}
In the theory of optical gap solitons, slowly-moving
finite-amplitude Lorentzian solutions are
found to mediate the transition from bright to coexistent dark-antidark
solitary wave pairs when the laser frequency is detuned out of the
proper edge of a dynamical photonic bandgap. Catastrophe theory is applied
to give a
geometrical description of this strongly asymmetrical 'morphing' process.
\end{abstract}
\pacs{42.65.-k,42.65.Tg,42.70.Qs}

The confinement of optical radiation in periodic media (gratings) with
nonlinear response occurs in the form of gap solitons
(GS), or more properly solitary waves,
as first predicted by Chen and Mills \cite{ChenMills},
and studied extensively afterwards
\cite{ale89,christo89,fk93,gaprev,millsbook,fiberexp,AsGa}.
In Kerr media, the prototype model for
GS is the following system of hyperbolic
PDEs with Hamiltonian (conservative) structure
\cite{bara98,derossi98},
which couples the forward
$u_+(z,t)$  and backward $u_-(z,t)$ propagating envelopes at
Bragg carrier frequency $\omega_B$
\begin{eqnarray}
i\left( \partial_{t}+\partial_{z} \right) u_{+} + u_{-} +
\left( X\ \left| u_{-}\right| ^{2}+
S\ \left| u_{+}\right| ^{2}\right) u_{+}
&=&0,\nonumber \\
\label{PDE1} \\
i\left( \partial_{t}-\partial_{z} \right) u_{-} + u_{+}
+\left( X\ \left| u_{+}\right| ^{2}+
S\ \left| u_{-}\right| ^{2}\right) u_{-}
&=&0.
\nonumber
\end{eqnarray}%
Equations (1) have been conveniently written in usual
dimensionless units $z=\Gamma Z$ and $t= \Gamma V_B T$, where
$Z$ and $T$ are the real-world propagation distance and time,
$\Gamma$ is the Bragg coupling coefficient, and $V_B$ is the
group-velocity at Bragg frequency. Moreover $S$ and $X$
are coefficients which specify the relative weight of self-
and cross-phase modulation, and $u_{\pm}$ are
proportional to real-world envelope amplitudes.
Equations~(1) are usually analyzed with $S=1$
and $X=2$ (one of the two coefficients can be always
set to have unitary modulus by a suitable rescaling of the field amplitudes)
which describes {\em scalar} mode coupling,
e.g., in optical fiber gratings \cite{ale89,christo89,fk93,gaprev,millsbook}.
Conversely we find convenient to
leave them as generic coefficients in order to describe both the
cases of focusing ($S,X>0$ \cite{ale89,christo89,fk93,gaprev}) and defocusing
($S,X<0$ \cite{ChenMills}) nonlinearity,  as well as the two limit cases
$X=0$ and $S=0$, which arise, e.g., when the cubic nonlinearity
originates from cascading in quadratic media
\cite{trillo96,conti97}. Cascading adds improved flexibility
since it permits to control the sign of the effective Kerr nonlinearity
by tuning the wavevector mismatch.
We also emphasize that, in the case $S=0$, Eqs.~(1) reduce to
the integrable (by means of the inverse scattering method)
massive Thirring model, and
hence the localized waves are strictly speaking solitons.

Despite the fact that GS arise in a
variety of physical settings and models, the importance of
Eqs.~(1) is threefold: (i) they describe with reasonable
accuracy optical GS experimentally investigated
in fiber Bragg gratings \cite{fiberexp} and in corrugated
GaAs waveguides \cite{AsGa}; (ii) though not integrable by
inverse scattering method (except for the massive Thirring case),
the model allows to construct the whole family of solitary waves;
(iii) Eqs.~(1) have allowed to assess the occurrence of peculiar effects
such as the onset of oscillatory instabilities
\cite{bara98,derossi98}, ultimately related with
the absence of material dispersion
(i.e., second-order derivatives) which distinguishes Eqs.~(1) from
other soliton-bearing dispersive models (e.g., those
of the nonlinear Schr\"{o}dinger type).
GS of Eqs.~(1) have been studied for more than a decade,
and both bright \cite{ale89,christo89}
and dark \cite{fk93} solutions were reported.
Yet, the existence of such GS solutions and their bifurcations
(how they change qualitatively against changes of parameters)
were never investigated to full extent.
Here we unveil the bifurcation structure of GS,
restricting ourselves to subluminal solutions for physical reasons.
We show that moving Lorenztian GS mark the transition between
in-gap bright GS and dark-antidark GS
pairs which coexist either below or above
(depending on the focusing or defocusing
nature of the nonlinearity, respectively) the
edge of a suitably defined dynamical gap.

Following the notation of Ref.~\cite{derossi98},
the entire family of solitary waves of Eqs.~(1)
can be characterized by seeking solutions in the following form
\begin{eqnarray}
u_{+}(z,t) &=&U_{+} \sqrt{\eta(\zeta)}
~\exp\{-i\Delta t + i\left[ \beta \zeta
+\phi_{+}(\zeta) \right] \}\nonumber \\
\label{ansatz1} \\
u_{-}(z,t) &=&U_{-} \sqrt{\eta(\zeta)}
~\exp\{-i\Delta t + i\left[ \beta \zeta
+\phi_{-}(\zeta) \right] \}\nonumber
\end{eqnarray}
where $\beta \equiv \gamma v \Delta$ plays the role of GS propagation constant,
and the intensity $\eta$ and the chirp (nonlinear phase)
$\phi_{\pm}$ profiles depend on $\zeta \equiv \gamma (z-v t)$,
with $\gamma =(1-v^{2})^{-1/2}$ being the (subluminal, $|v|<1$) Lorentz factor.
Furthermore $U_{+}=A \sqrt[4]{\frac{1+v}{1-v}}$ and
$U_{-}=-s A \sqrt[4]{\frac{1-v}{1+v}}$
account for the velocity induced asymmetry between
the foward and backward modes, whereas
$s \equiv {\rm sign}\left[ X (1-v^{2}) + S (1+v^{2}) \right]$
is the sign of the nonlinearity which appears only through
the common overall coefficient $A=\gamma^{-1}
\left| 2 X\left( 1-v^{2}\right) + 2 S\left( 1+v^{2}\right) \right|^{-1/2}$.
Importantly the entire family of GS is characterized by two
independent parameters, namely the normalized
detuning $\Delta$ and soliton velocity $v$
(corresponding, in real-world units,
to $\delta \omega=\Gamma V_B \Delta$
and $V=V_B v$), which are related to
rotational and translational group symmetries of Eqs.~(1),
respectively \cite{derossi98}. Note
that $\Delta=0$ corresponds to the Bragg frequency,
and $v=0$ yields still GS, i.e. confinement of light
with zero velocity in the lab frame.

By direct substitution of Eqs.~(\ref{ansatz1}) into Eqs.~(1),
it is readily verified that the intensity $\eta$ and overall phase
$\theta \equiv \phi_{+}-\phi_{-}$
obey the following one-dimensional (thus
integrable) Hamiltonian system of equations
\begin{eqnarray}
\dot{\eta}&=&2\eta \sin \theta =
-\frac{\partial H}{\partial \theta },\nonumber \\
\label{ODE1} \\
\dot{\theta} &=&2\delta +2\cos \theta -\eta =
\frac{\partial H}{\partial \eta }, \nonumber
\end{eqnarray}%
where the dot stands for $d/d \zeta$,
$H=H(\eta,\theta)=2\eta \cos \theta +2\delta \eta -\eta^{2}/2$
is the reduced conserved ($\dot{H}=0$) Hamiltonian,
which now depends on the single parameter $\delta \equiv \gamma \Delta$.
In Eq.~(1) and hereafter we implicitly
assume to deal with the self-defocusing nonlinearity $s=-1$
originally considered in Ref.~\cite{ChenMills}, including the
case of vanishing $S$ or $X$.
The results can be readily extended to
the self-focusing case $s=1$ with the substitution
$\theta \rightarrow 2\pi -\theta$ and
$\delta \rightarrow -\delta$.
Importantly, the latter condition means that
the role of frequencies below and above Bragg frequency
must be simply interchanged.

The reduced system~(\ref{ODE1}) permits
to find the solitary waves of Eqs.~(1)
as the separatrix trajectories which are
homoclinic to (i.e., emanate from and return to) the
unstable fixed points $\eta_s, \theta_s$ of Eqs.~(\ref{ODE1}).
Given the constraint $\eta>0$ in Eq.~(\ref{ansatz1}),
these are easily found to be of two kinds:
(i) $\left( \eta_s ,\cos \theta_s \right) =
\left( 0,-\delta \right)$ for $\delta^{2}<1$,
which is associated with solutions
of the bright type since $\eta(\zeta=\pm \infty)=\eta_s=0$;
(ii) $\left( \eta_s,\theta_s \right) =\left( 2\left( \delta -1\right), \pi
\right)$  for $\delta>1$, which describe GS with non-vanishing background
or pedestal $\eta(\zeta= \pm \infty) =\eta_s=2\left( \delta -1\right)$.
For any fixed value of the parameter $\delta$,
the solitary waves correspond, in both cases,
to level curves of the Hamiltonian $H_s=H(\eta_s,\theta_s)$.

Let us first clarify the relation between the existence domain
of these two families of GS and the stopband or
forbidden gap of frequencies exhibited
by the grating in the linear operation regime.
Bright solutions exist for $|\delta|<1$, which is mapped
into the inner domain $\Delta^{2}+v^{2}<1$
of the parameter plane $\Delta,v$
(such representation was already adopted in
Refs.~\cite{derossi98,jena97}).
This unitary circle can be regarded
as a {\em dynamical gap}, since in this domain the linear problem
(i.e., Eqs.~(1) with $S,X=0$)
yields exponentially damped traveling-wave solutions
($u^{\pm} \propto \exp(-i\Delta t + i Q \zeta)$, with $Q^2<0$).
On the other hand, solutions with nonzero pedestal
exist only outside this dynamical gap
or unitary circle.

To clarify further the role of the soliton velocity $v$,
it is important to link the dynamical gap $\Delta^{2}+v^{2}<1$
with the bandgap (henceforth termed {\em rest gap} to rule out any
possible source of misunderstanding) of the stationary
linear coupled-mode problem.
Such rest gap is well known to be
related to the reflectivity bandwidth of the grating,
in turn measurable by means of a
tunable cw laser in the laboratory (or rest) frame (z,t).
Quantitatively, the rest gap is given by the frequency
range $|\delta_1|<1$ ($|\delta \omega_1|<\Gamma V_B$
in real-world units) where solutions
$u_{\pm}(z,t)=u_{\pm}(z) \exp(-i \delta_1 t)$
with frequency detuning $\delta_1$
(real-world frequency $\omega_B+\delta \omega_1$)
are exponentially damped in
space,  at variance with the out-gap case $|\delta_1|>1$
where they become oscillatory \cite{yariv}.
In the velocity-frequency plane ($v,\delta_1$)
the rest gap is the ($v$-independent) lighter shaded
domain shown in Fig.~1(a), which has a clear one-to-one
correspondence with the reflectivity bandwidth
reported for comparison in Fig.~1(b).
To map the dynamical bandgap in the same parameter plane,
we need to know what is the actual normalized GS
frequency detuning in the rest frame,
which is readily found to be $\delta_1 \equiv \gamma^2 \Delta$
by grouping in Eqs.~(2) phase terms proportional to $t$.
In other words the excitation of a GS characterized by the
parameters $v$ and $\Delta$ requires a source with detuning
$\delta_1=\gamma^2 \Delta$ from Bragg frequency.
As a consequence the dynamical gap where bright GS exist,
can be mapped onto the whole shaded domain
$\delta_1^2 < \gamma^4 (1-v^2)$ or equivalently $|\delta_1|<\gamma$
of Fig.~1. It is clear that, for any given velocity $v$,
this entails a frequency range which is wider than the
rest gap $|\delta_1|<1$ and reduces to it only
in the $v=0$ limit.
As a consequence bright GS exist for frequencies ranging from the rest gap
(reflectivity bandwidth) in the $v=0$
case, to the whole frequency axis as $|v| \rightarrow 1$
(i.e., as the soliton velocity $V$ approaches the linear group-velocity
$V_B$ in the forward or backward direction).
\begin{figure}
\leavevmode\setlength{\epsfxsize}{8cm}
\hspace*{-0.5cm} \centerline{\mbox{\epsffile{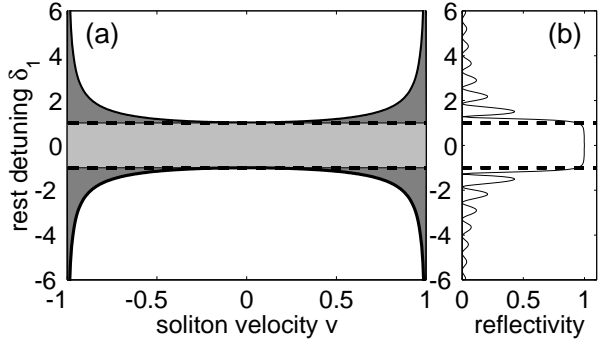}}}
\caption{(a) The dynamical photonic bandgap (whole shaded
domain) $|\delta_1|<\gamma$ in the parameter plane of
velocity and rest detuning ($v,\delta_1$).
Such domain is mapped back onto the inner domain bounded
by the circle $\Delta^2+v^2=1$
in the plane ($\Delta,v$), see Fig.~2.
The rest bandgap $|\delta_1|<1$ is the
smaller region between the two dashed lines $\delta_1=\pm 1$
(light shaded area),
and corresponds to the bandwidth of the linear reflectivity
curve shown with the same vertical scale in (b)
for a grating of normalized length
$z_L=\Gamma L=4$.}
\label{gap}
\end{figure}
As far as the terminology is concerned, a last important
comment is in order.
Bright solitary solutions of Eqs.~(1)
are usually classified as gap solitons or Bragg solitons,
depending on their detuning $\delta_1$ being inside
($|\delta_1|<1$, light shaded domain in Fig. 1) or outside
($1<|\delta_1|<\gamma$, dark shaded domain in Fig. 1)
the rest gap, respectively. Though such a distinction
can be useful to locate the operating frequency with respect to the
reflectivity bandwidth of the grating, it appears otherwise rather arbitrary.
First, for any given velocity $v$ there are no qualitative changes of the
solutions by crossing the boundary $|\delta_1|=1$ between these two regions.
Second and more important, the picture of Fig.~1 suggests that
all the existing bright localized waves are in fact {\em gap} solitons if
referred to the dynamical gap. In other words the effective
frequency gap seen by a soliton which moves at velocity $v$
is wider (the faster the soliton the wider the gap) than the rest gap
measured through the reflectivity.  This is also supported by the fact that
$\delta$ can be interpreted as the frequency detuning of the soliton
in Lorentz transformed variables $\zeta,\tau$ with $\tau=\gamma (t-vz)$
\cite{comment}. Therefore the rest gap condition $|\delta_1|<1$ must be
replaced
in moving soliton coordinates by the condition
$|\delta|=|\delta_1|/\gamma<1$  which coincides with the
dynamical gap.

Equations (\ref{ODE1}) are equivalent to the motion of an ideal particle
of unitary mass and total energy $E$ in a quartic potential well $U(\eta)$,
ruled by the equation $\ddot{\eta}=-\partial U/\partial \eta$.
The kinetic energy is easily obtained in the standard form
from the first of Eqs.~(\ref{ODE1}) by eliminating
$\sin \theta$ through $H$, which yields
\begin{equation}
\dot{\eta}=\sqrt{2\left[E-U(\eta)\right]},
\label{kinetic}\end{equation}
where $U=-2\delta H\eta -\frac{1}{2}\left( 4-4\delta^{2}-H\right) \eta
^{2}-\delta \eta^{3}+\frac{1}{8}\eta^{4}$, and $E=-H^2/2$.
The GS solutions can be worked out explicitly by inverting
the quadrature integral obtained from Eq.~(\ref{kinetic})
with the energy $E=-H_s^2/2$ pertaining
to the unstable fixed point.
We found the following expressions for the case (i) $|\delta|<1$,
entailing bright GS
\begin{eqnarray}
\eta_{B}&=&\frac{4\left( 1-\delta ^{2}\right) }{\cosh \left(
2\sqrt{1-\delta ^{2}}\zeta \right) -\delta },\label{brightint} \\
\theta_B&=&\tan^{-1}
\left[\frac{\sqrt{1-\delta^2}\,\sinh(2\sqrt{1-\delta^2}~\zeta)}{-1+\delta\,
\cosh(2
\sqrt{1-\delta^2}~\zeta)}\right].
\end{eqnarray}
Close to the low-frequency edge of the dynamical gap $\delta \sim -1$,
the following approximation of the intensity profile in Eq.~(\ref{brightint})
holds valid [exploit $\cosh(2x)=2\cosh^2(x)-1$
and $1-\delta^2 \sim 2(1+\delta)$]
\begin{equation}
\eta_{B}=4(1+\delta)~{\rm sech}^2\left(\sqrt{2(1+\delta)}\zeta \right),
\label{NLS}\end{equation}
which is characteristic of the one-soliton solution
of the focusing nonlinear Schr\"{o}dinger
equation, which provides a reasonable description of GS
in this region of the gap \cite{gaprev,nlsrecent}.

In the case (ii), i.e. for $\delta>1$ two
solutions coexist, being associated to two branches of a double-loop
separatrix.
The first one represents a dark soliton
\begin{eqnarray}
\eta_{DK}&=& 2\left( \delta -1\right) \frac{\cosh \left(
2\sqrt{\delta -1}\zeta \right) -\sqrt{\delta }}{\sqrt{\delta
}\cosh \left( 2\sqrt{\delta -1}\zeta \right) +1},\\
\theta_{DK}&=&\tan^{-1}
\left[\frac{4\sqrt{\delta-1}(\delta+1)\sinh(2\sqrt{\delta-1}\zeta)}
{3\delta^2+2\delta^\frac{3}{2}-3+4(\delta-1)\delta\,\cosh(2\sqrt{\delta-1}\zeta)
+
(\delta^2-2\delta^\frac{3}{2}-1)\cosh(4\sqrt{\delta-1}\zeta)}\right],
\end{eqnarray}
whereas the second one is a bright on pedestal
or so-called antidark solution \cite{antidark}
\begin{eqnarray}
\eta _{AK}&=&2\left( \delta -1\right)
\frac{\sqrt{\delta }\cosh \left( 2\sqrt{\delta -1}\zeta \right)
+1}{\sqrt{\delta }\cosh
\left( 2\sqrt{\delta -1}\zeta \right) -1},\\
\theta_{AK}&=&\tan^{-1}
\left[ \frac{4\sqrt{\delta-1}\sqrt{\delta}\sinh(2\sqrt{\delta-1}\zeta)}
{2-3\delta+\delta\cosh(4\sqrt{\delta-1}\zeta)} \right].
\end{eqnarray}
These dark and antidark solutions specialized
to the zero-velocity case ($v=0$) have important implications
in terms of the stationary ($\partial_t=0$) response of the grating,
inducing limiting or frustrated bistability, as discussed
in Ref.~\cite{chaos}.

Right on the high-frequency edge of the dynamical gap,
i.e. for $\delta \rightarrow 1$,
the dark GS vanishes ($\eta_{DK}\rightarrow 0$),
while the bright and the antidark
solutions have the following common limit
\begin{eqnarray}
\eta_{LZ}&=&\frac{8}{1+\left( 2 \zeta \right)^{2}},\\
\theta_{LZ}&=&\tan^{-1} \left[ \frac{4\zeta}{\left(2\zeta\right)^2-1} \right],
\label{LZfa}\end{eqnarray}
which represents a finite-amplitude
moving Lorentzian soliton, i.e., a GS
with Lorentzian intensity profile.
The existence of an exact solution of Eqs.~(1) with
non-exponentially decaying tails can be
understood from the dynamical system (\ref{ODE1}) as being associated with
a degenerate fixed point at the origin with zero eigenvalues.
The fact that the Lorentzian shape approximates well bright GS
close to the upper bound of the gap was noticed earlier
under strictly stationary conditions ($v=0$) \cite{millsbook}.

Viceversa, for $\delta \rightarrow -1$,
i.e., close to the low-frequency edge of the dynamical
gap, the intensity of the bright GS in Eq.~(\ref{brightint})
[or, consistently, of its nonlinear Schr\"{o}dinger approximation (\ref{NLS})]
reduces to the following Lorentzian profile
\begin{equation}
\eta_{LZ}=\frac{4(1+\delta)}
{1+4\frac{1+\delta}{1-\delta}\zeta^2}\cong
\frac{4(1+\delta)}{1+2(1+\delta)\zeta^2}.
\label{zeroLZ}\end{equation}
In this case, however, as the stopband edge is
approached the Lorentzian GS (\ref{zeroLZ})
becomes broader and smaller,
and eventually vanishes identically ($\eta_{B} \rightarrow 0$)
for $\delta=-1$.
The difference between the finite-amplitude [Eq.(\ref{LZfa})]
and the vanishing-amplitude [Eq.(\ref{zeroLZ})]
Lorentzian GS accounts for an intrinsic asymmetry
of the nonlinear grating operation with
respect to interchange of frequencies below and above
the Bragg frequency, respectively.
\begin{figure}
\leavevmode\setlength{\epsfxsize}{8cm}
\hspace*{-0.5cm} \centerline{\mbox{\epsffile{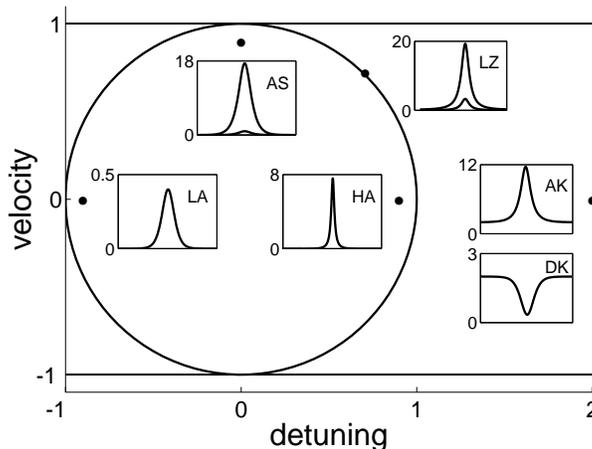}}}
\caption{Existence diagram (defocusing case) for subluminal GS in
the region of detuning-velocity $(\Delta,v)$ plane
bounded by the thin lines at $v=\pm 1$. The insets show GS intensity profiles
$|u_{\pm}|^2$ sampled at the  nearest marker (filled circle). Bright GS of
low-amplitude
(LA), high-amplitude (HA), and asymmetric (AS) types exist
inside the dynamical gap (unitary circle $\Delta^2+v^2=1$). HA solitons become
finite-amplitude Lorentzian (LZ) over the right semi-circle, and then
bifurcate into dark (DK) and antidark (AK) pairs
which coexist for frequencies above the upper edge
of the dynamical bandgap.
In the focusing case the same picture holds with $\Delta \rightarrow -\Delta$.}
\label{circlefig}
\end{figure}
Figure~\ref{circlefig} summarizes the nature of the different
GS solutions by reporting the intensity profiles $|u_{\pm}|^2$
of qualitatively different GS solutions sampled in the parameter plane
($\Delta,v$).  First, it must be noticed that
still ($v=0$) GS have equal intensities $|u_+|^2=|u_-|^2$
as a consequence of the fact that the net photon
flux is zero, for both in-gap (bright) and out-gap (dark-antidark) solutions.
Moreover, let us recall that the amplitude of the
in-gap (bright) GS increases and their width decreases
by spanning the gap from left to right.
The symmetry $|u_+|^2=|u_-|^2$ is broken for moving GS, which
have a stronger component in the direction of motion (i.e., $|u_+|>|u_-|$
for $v>0$
and $|u_-|>|u_+|$ for $v<0$).
Importantly, bright GS which have high-amplitude
(see HA inset in Fig.~\ref{circlefig})
close to the upper edge of the dynamical gap,
become Lorentzian GS (LZ inset in Fig.~\ref{circlefig})
over the edge $\Delta+v^2=1$, and then bifurcate into
dark-antidark pairs (DK and AK insets in Fig.~\ref{circlefig}) outside the
dynamical
bandgap.  Viceversa, as explained above,
the low-amplitude (LA inset)
GS which exist close to the lower edge of the gap
vanishes in the limit $\delta \rightarrow -1$,
and no solutions exist below the bottom
of the stopband (i.e., outside the circle for $\Delta<0$).
For a focusing nonlinearity
($s=1$) an identical picture with $\Delta \rightarrow -\Delta$ holds true,
meaning that the dark-antidark pairs originate always from
the high-amplitude bright GS, though, in this case, they now
exist below the low-frequency edge of the dynamical gap.

\begin{figure}
\leavevmode\setlength{\epsfxsize}{6.5cm}
\hspace*{-0.5cm} \centerline{\mbox{\epsffile{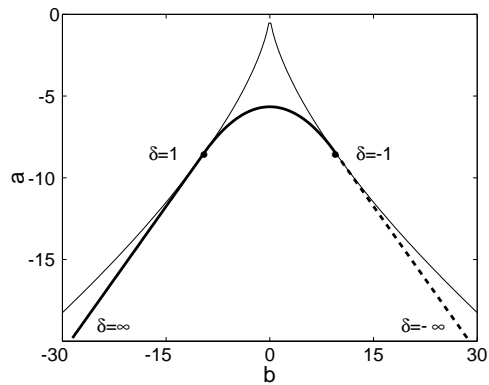}}}
\caption{Plot of the bifurcation set (thin line)
from Eq.~(\ref{bifset}) and the control line (thick line)
which gives the parametric dependence of the coefficients
$a$ and $b$ of the potential $\hat{U}(x)$ in Eq.~(\ref{standardpot})
on the control parameter $\delta$, when this is varied
from large negative values ($\delta=-\infty$)
to large positive values ($\delta=\infty$).
The dashed portion of the control line correspond to
unphysical solutions ($\eta<0$).
Two catastrophes occur at the points $\delta=\pm 1$.}
\label{cuspfig}
\end{figure}

The bifurcation of GS can be effectively explained in term of the
catastrophe theory \cite{gilmore}, and the underlying
classification of the singularities of smooth functions.
The quartic potential $U(\eta)$ belongs to the so called cusp $A_{+3}$
family \cite{gilmore}. It can be recast in the
following standard form by means of the change of variable
$\eta =\sqrt[4]{2}x + 2\delta$,
thus obtaining
\begin{equation}
\hat{U}(x)=\frac{1}{4}x^{4}+\frac{a}{2}x^{2}+bx,
\label{standardpot}
\end{equation}
where $a \equiv \sqrt{2}\left( -4-2\delta ^{2}+H\right)$
and $b \equiv -8\sqrt[4]{2}\delta$.
A necessary condition for the solitary waves to exist
is that the potential $\hat{U}(x)$ has
three critical points (i.e., such that $\partial \hat{U}/\partial x =0$),
thus being of a double-well type.
In the parameter plane $(a,b)$ this occurs in a domain
bounded by the following curve, so-called ''bifurcation set'' \cite{gilmore},
\begin{equation}
\left( \frac{a}{3}\right) ^{3}+\left( \frac{b}{2}\right)^{2}=0,
\label{bifset}\end{equation}
which is reported as a thin solid line in Fig.~\ref{cuspfig}.
As shown this curve has the characteristic shape of a cusp.
In the spirit of the catastrophe theory, we also report
in Fig.~\ref{cuspfig} the so-called control line, i.e.,
how the parameters $(a,b)$, and as a consequence
the potential $\hat{U}$, vary by changing the single control
parameter $\delta$ from large negative values to large positive values
(indicated in Fig.~\ref{cuspfig} by the limit
$\delta=-\infty$ and $\delta=\infty$,
respectively). In doing so we calculate $a$ with $H=H_s$, i.e. the value
of the Hamiltonian pertaining to the solitons.
According to our analysis, by varying $\delta$ in the range
$(-\infty,\infty)$ the control line remains always inside the cusp,
indicating the possibility to have solitary waves.
Dramatic qualitative changes of the solutions must be expected at the
catastrophe points $\delta=\pm 1$ where the control line
becomes tangent to the bifurcation set. In spite of the
apparent symmetry of Fig.~\ref{cuspfig}, the two catastrophes
are the signature of the strongly asymmetrical behavior of GS
against interchange of frequencies below ($\delta<0$)
and above ($\delta>0$) the Bragg frequency.
Indeed, the $\delta=1$ catastrophe marks the point where
the maximum of the potential (\ref{standardpot})
moves from a finite positive value
for $\delta>1$ (which makes accessible two distinct asymptotic
evolutions inside the two wells, in turn corresponding to the
dark and antidark GS), to the origin for $-1<\delta<1$
(where only one well is accessible for the asymptotic motion towards
the origin which describes the bright GS).
Conversely, the other catastrophe at $\delta=-1$ marks
the point where these bright GS simply cease to exist
because the maximum of the potential moves towards negative values.
In fact, though the potential still has a double-well
shape even for $-\infty<\delta<-1$,
the possibility to have solitary waves is ruled out
by the fact that the two wells become accessible only with $\eta<0$,
and hence the solutions are unphysical (recall that $\eta$ is an intensity).
The dashed line in Fig.~\ref{cuspfig} displays
that portion of the control line where the solutions
are unphysical.

In summary we have shown that GS solutions of a well-known
standard coupled-mode model with
Kerr or Kerr-equivalent nonlinearity undergo a bifurcation
which is strongly asymmetrical with respect to the Bragg frequency.
The qualitative change of the solutions is explained with
a geometrical picture based on the application
of the catastrophe theory. The bifurcation is marked by
the existence of finite-amplitude Lorentzian GS.
In this sense it is reminiscent of the recently investigated
case of localized waves sustained by a gap of full nonlinear origin
\cite{ngs}.
In spite of the diversity between the bifurcation discussed
here and that of Ref.~\cite{ngs}, this suggests that Lorentzian solitons
can play a universal role in the localization of light in periodic media.

\acknowledgements
We thank Yuri Kivshar for fruitful discussions concerning Lorentzian solitons.



\begin{references}

\bibitem{ChenMills} W. Chen and D. L. Mills,
Phys. Rev. Lett. {\bf 58}, 160 (1987).

\bibitem{ale89} A. B. Aceves and S. Wabnitz,
Phys. Lett. A {\bf 141}, 37 (1989).

\bibitem{christo89} D. N. Christodoulides and R. I. Joseph,
Phys. Rev. Lett. {\bf 62}, 1746 (1989).

\bibitem{fk93} J. Feng and F. K. Kneub\"{u}l, J. Quantum
Electron. {\bf QE-29}, 590 (1993).

\bibitem{gaprev} C. M. De Sterke and J. E. Sipe,
in {\em Progress in Optics XXXIII}, edited by E. Wolf,
(Elsevier, Amsterdam, 1994), Chap. III.

\bibitem{millsbook}
D. L. Mills, {\it Nonlinear optics} (Springer, New York, 1998).

\bibitem{fiberexp}
B. J. Eggleton, R. E. Slusher, C. M. de Sterke, P. A. Krug, and J. E.
Sipe, Phys. Rev. Lett. {\bf 76}, 1627 (1996);
B. J. Eggleton, C. M. de Sterke, R. E. Slusher,
J. Opt. Soc. Am. B {\bf 14}, 2980 (1997).

\bibitem{AsGa}
P. Millar, R. M. De La Rue, T. F. Krauss,
J. S. Aitchison, N. G. R. Broderick, and  D.J. Richardson,
Opt. Lett. {\bf 24}, 685 (1999).

\bibitem{bara98} V.I. Barashenkov, D.E. Pelinovsky, and E.V.
Zemlyanaya, Phys. Rev. Lett. {\bf 80}, 5117 (1998).

\bibitem{derossi98} A. De Rossi, C.
Conti, and S. Trillo, Phys. Rev. Lett. {\bf 81}, 85 (1998);
Opt. Lett. {\bf 23}, 1265 (1998).

\bibitem{trillo96} S. Trillo,
Opt. Lett. {\bf 21}, 1732 (1996).

\bibitem{conti97}	C. Conti, G. Assanto, and S. Trillo,
Opt. Lett. {\bf 22}, 1350 (1997).

\bibitem{jena97}  T. Peschel, U. Peschel, F. Lederer,
and B. A. Malomed, Phys. Rev. E {\bf 55}, 4730 (1997).

\bibitem{yariv} A. Yariv, {\it Optical electronics in modern
telecommunications}, (Oxford University Press, New York, 1997).

\bibitem{comment} The Lorentz transformed time is written with
light velocity $c=1$, consistently with the normalization of Eqs.~(1).
Let us point out that Eqs.~(1) are not Lorentz-invariant
except in the integrable case $S=0$ of the massive Thirring model.
However, the consequence that, in general, moving GS cannot be generated
by means of a Lorentz transformation of the rest GS does not prevent
from using Lorentz transformed variables (see also Ref.~\cite{bara98}).

\bibitem{nlsrecent} C. M. de Sterke and B. J. Eggleton,
Phys. Rev. E {\bf 59}, 1267 (1999);
C. M. de Sterke, D.G. Salinas, and J.E. Sipe,
J. Opt. Soc. Am. B {\bf 16}, 587 (1999).

\bibitem{antidark} Y.S. Kivshar and V.V. Afananasjev,
Phys. Rev. A {\bf 44}, R1446 (1991).

\bibitem{chaos} S. Trillo, C. Conti, G. Assanto, and A. V. Buryak,
Chaos {\bf 10}, 590 (2000).

\bibitem{gilmore} R. Gilmore, {\it Catastrophe theory for scientists and
engineers} (Dover Publ., New York, 1993).

\bibitem{ngs} C. Conti, S. Trillo, and G. Assanto,
Phys. Rev. Lett. {\bf 85}, 2502 (2000).

\end{references}
\end{document}